\documentstyle[12pt,epsf]{article}

\begin{document}
\vspace*{.15in}
\noindent

\baselineskip 1.7pc
\begin{center}
{\Large \bf Cluster Model Analysis of Kaon Scattering from $^{12}$C}

\vspace{.5cm} {\large  Ahmed A. Ebrahim} \\
{Physics Department, Assiut University,
Assiut 71516, Egypt}
\end{center}
\date{}
\vskip .5 cm
\begin{abstract}
Angular distributions of differential cross sections for the
interaction of $K^{+}$ mesons with $^{^{12}}$C nucleus at beam
momenta of 635, 715, and 800 MeV/c have been analyzed using
3$\alpha$-particle model of  $^{^{12}}$C.   Differential cross
sections for inelastic transitions to the (2$^{^{+}}$;
 4.44 MeV) and (3$^{-}$; 9.64 MeV) states in $^{^{12}}$C are calculated, and deformation lengths
$\delta_{2}$ and $\delta_{3}$ are extracted and consistent with
 other works. Good
agreement with experimental data of elastic and inelastic
$K^{\pm}$-$^{12}$C scattering is obtained.

\end{abstract}

\section{Introduction}
The $K^{+}$nucleus scattering is of considerable interest because of
relative weak kaon nucleon ($KN$) interaction. $K^{+}$-nucleus
 cross sections should be fairly weak and the amplitudes reasonably simple because of the simplicity of $K^{+}N$ system. The
 $K^{+}$-nucleus
 interaction
  is the weakest of any strongly interacting probe, and the resulting mean free path is expected to be large (in the order of 7 {\em fm}).
   Therefore,
$K^{+}$ can penetrate quite freely into the central region of the
target nucleus, having time to undergo predominantly single $K^{+}N$
collisions and appearing to be a good probe for the distribution of
neutrons in nuclei, while electron scattering is applied to study
the nuclear charge density. There is no true
 absorption of $K^{+}$ to complicate things, in distinction to the case of low energy $\pi$-nucleus scattering, where the mean free path
  is also large. Another difference is that $K^{+}$ wave lengths are small enough to sense
   individual nucleons within nuclei, where $\lambda$=2 {\em fm} for 50 MeV pion kentic energy while $\lambda$=0.5 {\em fm} for 715 MeV/c kaon
    lab momentum \cite{CBD}.

On the contrary, the $K^{-}N$ system has many narrow resonances
($\Gamma\simeq$ 15-40 MeV) and there are open channels down to the
$K^{-}N$ threshold.
 Cross sections are comparable to those for NN scattering. Therefore, we expect that $K^{-}$-nucleus cross sections reflect
  these features: The interaction is strong and the amplitudes are expected to be complicated \cite{MAR}.

An earlier elastic scattering experiment at 800 MeV/c \cite{MAR} has
gained interest since its results are higher than expected by
optical potential model calculations. This led to the concept of
'swollen nucleons', with the elementary
  projectile-nucleon coupling within the nucleus greater than that found in free space. Agreement in magnitude is found for an enhancement of
   the fundamental cross section by 33\% \cite{CMC}. Since, however, the normalization uncertainty of that elastic scattering
    experiment was about 19\%, some of this enhancement of the theory may not be necessary. Total cross sections for
$K^{+}$ mesons on several nuclei at a range of kaon lab momenta also
exceeded model expectations \cite{MFJ,SIE}, indicating that nucleons
within nuclear medium do not behave as they do in free space. These
are analyzed by Siegel {\it et al.} \cite{SIE} with the conclusion
that agreement with theory is possible. In addition,
 there may be errors in the $KN$ phase shifts used, although the agreement with the $K^{+}$-$^{40}$Ca results indicate these are not major.
  By comparing $K^{+}$-$^{12}$C results with $K^{+}$-deuterium scattering, the effects of such errors are reduced. Siegel {\it et al.} \cite{SIE}
  suggest that within the nucleus, there is an increase of the S11 $KN$ phase shift than that given by Martin
    \cite{MAR}. Two explanations have been  offered. Siegel {\it et al.} suggest that the effect arises from an increase of nucleon size.
     Brown {\it et al.} \cite{GEB} ascribe this increase due to change
      in mass of the $\rho$ and $\omega$ which mediate the $K^{+}$ reaction in the nuclear medium. This leads to an optical potential that
       depends nonlinearly on the nucleon density giving rise to an  increased repulsion (over the first-order $t\rho$) and a decreased effective
        nuclear radius. The agreement with experiment can be obtained by changing the mass of $\rho$ and $\omega$ in the nuclear medium.

The  kaon-nucleus optical potential is based on the $K^{+}N$
t-matrix and the nuclear one particle wave functions of a square
well potential \cite{PBS}. The agreement with experimental data was
obtained with increasing the S11  $K^{+}N$ phase shift by 10-15\%
which is taken as in-medium effect. Also,  study of $K^{+}$
scattering with a local version \cite{AAE} of Kisslinger potential
\cite{LSK} is applied. This has been early suggested for
pion-nucleus scattering.  A 6-parameter Woods-Saxon
  potential
is fitted to experimental data \cite{AAB}. However, ambiguities
appear in the values of these parameters when adjusted to
 experimental data. When inelastic calculations are forced to give the known transition strengths, the ambiguity in parameters can be removed.

Recently, coupled-channels calculations for elastic and inelastic
scattering of $K^{+}$ at 715 MeV/c by $^{6}$Li and $^{12}$C at 635,
715, and 800 MeV/c kaon Lab
 momenta have been analysed \cite{EBR}, using the Watanabe superposition model in a Woods-Saxon shape, together with coupled-channels Born
  approximation. This was carried out in terms of alpha-particle and deuteron optical potentials, taking into account alpha-deuteron and
   3$\alpha$-clusters of $^{^{6}}$Li and $^{^{12}}$C nuclei. The elastic scattering studies
resulted in considerable ambiguities in WS parameters. Using best
fit parameters, calculations were carried out for inelastic
scattering of kaons to the lowest 2$^{+}$ and 3$^{-}$ states of
$^{12}$C and 3$^{+}$ of $^{6}$Li. The corresponding deformation
parameters are extracted and ambiguities in WS parameters are
greatly reduced. It is concluded that the Watanabe superposition
model in WS  potential form seems to be useful for kaon-nucleus
scattering.

In the present work the distorted-wave Born approximation (DWBA) is
applied for the interaction between the projectile and the whole
target nucleus. We derive an optical potential for the reaction
$K^{\pm}$-$^{12}$C when the nucleus
 is considered to be constructed of three alpha clusters together with the
DWUCK4 code \cite{PDK}. The derived optical potential is employed to  successfully predict the
angular distributions
 of the differential cross sections of the $K^{\pm}$ elastically and inelastically scattered to the lowest
 2$^{+}$ and  3$^{-}$ states in $^{12}$C in the momentum range of 635 to 800 MeV/c.
\section{The Folding Model}

 The DWUCK4 code which solves the nonrelativistic
Schr\"odinger equation, and was originally written to calculate the
scattering and reaction observables for binary nuclear reactions, is
used to calculate the angular distribution for kaon scattering from
a nucleus. Such a program should be provided with an effective kaon
mass, the target mass and an effective kaon energy. This requires
transforming the true kaon mass $m_{K}$ and beam energy $T_{K}$ to
surrogate values $E_{L}$ and $M_{K}$ \cite{LER}, using $M_{K}
=\gamma_{K}\, m_{K}$ with $m_{K}$ the free charged kaon mass of
0.530 amu or 493.707 MeV and
\begin{eqnarray}
\gamma_{K}=(x\,+\,\gamma_{L}) / (1 \,+ \,x^{2}\, +\,2\, x\, \gamma_{L})^{1/2},\qquad{ x\,=\, m_{K} /m_{T}},
\qquad{\gamma_{L}=1 + (T_{K}/m_{K}\, c^{2})}.\nonumber
\end{eqnarray}
Here, $m_{T}$=target mass= $A$ times the amu.
It is convenient to use the center of mass kinetic energy $E_{cm}=20.901\, k^{2}\, (M_{K}\, +\,m_{T}) / M_{K} m_{T}$  MeV, with
$M_{K}$ and $m_{T}$ in amu, with $k\,=\,(m_{K} c/\hbar)\,\, \sqrt{( \gamma_{K}^{2} \,-\,1)}$, giving
$E_{L}=E_{cm}\, (M_{K}\, +\, m_{T}) / m_{T}$.\\
As an example, for a $T_{K}$=310.638 MeV ($P_{lab}$=635 MeV/c) charged kaon incident upon $^{12}$C, one replaces the true kaon mass
into the  DWUCK4 code with $M_{K}$ =0.82852 amu = 771.6 MeV, and the actual beam energy with $E_{L}$=260.54 MeV.
A parameter-free version of the local potential method has been applied with good success to a
limited range of kaon-nucleus data \cite{AAE}.

$^{12}$C is considered to consist
 of three $\alpha$-particles; each $\alpha$-particle is bound much more weakly than a nucleon in the $^{12}$C nucleus.
  In the following, we derive an analytical expression of $K^{\pm}$-$^{12}$C potential with the 3$\alpha$-cluster model of
$^{12}$C.  The resulting potential is inserted into the DWUCK4 code
to calculate the differential cross sections. The calculations are
compared to experimental
 data \cite{MAR, REC}.

The total optical potential of $K^{\pm}-\alpha$ scattering can be described in the usual Woods-Saxon (WS) radial dependence
\begin{equation}
V_{\alpha}(R)\,=\,V_{C}(R)-\,U_{N}(R),
\end{equation}
where $R$ is the separation distance between the kaon and
$\alpha$-particle,
 $V_{C}$(R) is the Coulomb potential which is taken to be that of
a uniform charged sphere of radius $R_{C}$\,=\,1.2\,$A^{1/3}$ {\em
fm} \cite{BRA}, and $A$ is the target mass number. The nuclear
potential $U_{N}(R)$ is defined to be
\begin{eqnarray}
U_{N}(R)\,=\,V\,f(R,R_{v},a_{v})\,+\,i\,W\,f(R,R_{w},a_{w}),\nonumber
\end{eqnarray}
where
$f(R,R_{i},a_{i})\,=\,[1\,+\,\exp(R\,-\,R_{i})/a_{i}]^{-1}\,\,\,
(i=v,w)$ is the Woods-Saxon function.
 $V, R_{v}(=r_{v}\,A^{1/3})$, and
$a_{v}$ are, respectively, the depth, radius, and diffuseness of the
real part of the potential and $W, R_{w}(=r_{w}\,A^{1/3})$, and
$a_{w}$ are the corresponding quantities for the imaginary part.
Since kaons are spinless particles, the spin-orbit term is not taken
 into account.

To fit the experimental distributions, the optical model parameters were varied systematically to minimize the quantity:
\begin{equation}
\chi^{2}\,=\frac{1}{N}\,\sum_{i=1}^{N}\,\left({
\frac{{{\sigma_{exp}(\theta_{i})}}
\,-\,{{\sigma_{th}(\theta_{i})
}}}{\Delta\sigma_{exp}(\theta_{i})}}\right)^{2},
\end{equation}
where $N$ is the total number of the experimental points,
$\sigma_{th}(\theta_{i})$ is the predicted differential cross
section at angle $\theta_{i}$ and $\sigma_{exp}(\theta_{i})$ and
$\Delta\sigma_{exp}(\theta_{i})$ are the experimental cross section
and its associated error, respectively, where relative errors of
10\% \cite{BOY}-\cite{CLM} were taken for computing $\chi^{2}$.

In the framework of folding model, we can take the interaction between $K^{\pm}$ and $\alpha$-particle as the elementary interaction to fold
 the $\alpha$-particle density of the target nucleus $^{12}$C, and then obtain a single folding potential as
\begin{equation}
V_{^{12}C}(R)=\int \mid
\psi(r,\rho)\mid^{2}[V_{\alpha} (\mid
\bar{R}\,+\,\frac{2}{3}\,\bar{\rho}\mid)\,+\,V_{\alpha}(\mid\bar{R}\,-\,\frac{1}{3}\,\bar{\rho}\,
-\,\frac{1}{2}\,\bar{r}\mid)\,+\,V_{\alpha}(\mid\bar{R}\,-\,\frac{1}{3}\,\bar{\rho}\,
+\,\frac{1}{2}\,\bar{r}\mid)]\,d\bar{r}\,\,d\bar{\rho},
\end{equation}
where $\bar{R}$ is the separation vector between the centers of
the two colliding particles. The internal position vectors $\bar{r}$
and $\bar{\rho}$ are defined by the position vectors of the three
alpha particles $\bar{R}_{1}$, $\bar{R}_{2}$ and $\bar{R}_{3}$
constituting $^{^{12}}$C nucleus so that
\begin{eqnarray}
\bar{r}=\bar{R}_{2}\,-\,\bar{R}_{1} \,\,\,\,\,{\rm and}\,\,\,\,\,
\bar{\rho}=\bar{R}_{3}\,-\,\frac{1}{2}\,(\bar{R}_{1}\,+\,\bar{R}_{2})
,\nonumber
\end{eqnarray}
The nuclear matter density
distribution for $\alpha$-particle
\begin{equation}
\rho_{\alpha}(r)= \rho_{0}\,\,\exp(-\beta\,r^{2}),
\end{equation}
where $\rho_{0}$=0.4229 {\em fm$^{-}$} and $\beta$=0.0.7014 {\em fm$^{-2}$} \cite{GRA}.

The optical potential of each kaon-alpha in $^{^{12}}$C is given by
$V_{\alpha}$. The internal wave function of $^{12}$C has the form
\cite{MWK}
\begin{equation}
\psi(r, \rho)=(\frac{2\sqrt{3}\,\mu}{\pi})^{\frac{3}{2}}
\,\,\exp(-\mu\,(2\,\rho^{2}\,+\, \frac{3}{2}\,r^{2})),
\end{equation}
where $\mu$ is the range parameter of $^{12}$C discussed below in
section 3. We can decompose equation (3) into:
\begin{equation}
V_{^{12}C}(R)= V_{\alpha1}(R)+V_{\alpha2}(R)+V_{\alpha3}(R),
\end{equation}
where
\begin{eqnarray}
V_{\alpha1}(R)=\int \mid
\psi(r, \rho)\mid^{2}[V_{\alpha} \,(\mid
\bar{R}\,+\,\frac{2}{3}\,\bar{\rho}\mid)]\,d\bar{r}\,d\bar{\rho},\nonumber\\
V_{\alpha2}(R)=\int \mid
\psi(r, \rho)\mid^{2}[V_{\alpha}\,(\mid\bar{R}-\frac{1}{3}\,\bar{\rho}
-\frac{1}{2}\,\bar{r}\mid)]\,d\bar{r}\,d\bar{\rho},\nonumber\\
V_{\alpha3}(R)=\int \mid
\psi(r, \rho)\mid^{2}[V_{\alpha}\,(\mid\bar{R}-\frac{1}{3}\,\bar{\rho}
+\frac{1}{2}\,\bar{r}\mid)]\,d\bar{r}\,d\bar{\rho}.
\end{eqnarray}

Since $\psi(r,\rho)$ depends explicitly on the spatial coordinates
$r$ and $\rho$ and $\psi(r,\rho)$ is spatially symmetric about the
exchange of nucleons \cite{KEA}, the integral is not changed if the
carbon coordinate system is reoriented each time so that the
$\alpha$ cluster, which is being folded, lies along the vector
$\bar{\rho}$. This allows the potential to be written as
\begin{eqnarray}
V_{^{12}C}(R)&=& 3\,V_{\alpha1}(R).\nonumber\\
             &=& 3\,\int \mid
\psi(r, \rho)\mid^{2}\,V_{\alpha} \,(\mid
\bar{R}\,+\,\frac{2}{3}\,\bar{\rho}\mid)\,d\bar{r}\,d\bar{\rho}.
\end{eqnarray}

Replacing the variable $\bar{\rho}$ by $\bar{y}$ defined by
($\bar{y}$= $\bar{R}\,+\,\frac{2}{3}\,\bar{\rho}$) and taking
$\bar{R}$ as the polar axis for the angle integration with respect
to $\bar{y}$
\begin{equation}
V_{^{12}C}(R)=\,\frac{8\,\sqrt{\mu}}{\sqrt{\pi}\,R}\,\,e^{-\,9\,\mu\,R^{2}}\,
\int_{0}^{\infty}\,dy\,y\,\,e^{-\,9\,\mu\,y^{2}}\,\,V_{\alpha}(y)\,\,\sinh(18\mu
yR).
\end{equation}

The radial parts of the hadronic inelastic transition potential used here are:
\begin{equation}
F_{l}(R)= -\,\, \delta_{l}\hspace{1.mm}\frac{dV_{^{12}C}(R)}{dR},
\end{equation}
where $V_{^{12}C}$(R) is the optical potential found to fit the
corresponding elastic scattering (Eq. (9)). For a given transition,
we use $\delta_{l}$ to denote the corresponding real and imaginary
"deformation lengths" for the $K^{\pm}$ interaction, while $l$(=2 or
3) is the multipolarity; $\delta_{2}$ denotes the corresponding
deformation length for the transition to the (2$^{^{+}}$; 4.44 MeV)
and $\delta_{3}$ to the transition to the (3$^{-}$; 9.64 MeV)
 states in $^{12}$C.

\section{Results and Discussion}
The distorted wave Born approximation (DWBA) calculations are
performed using the computer code DWUCK4 \cite{PDK}. The radial
integrals have been carried out up to 40 {\em fm},  in 0.1 {\em fm}
steps with 100 partial waves. We analyzed the angular distribution
of differential cross sections for the elastic and inelastic
scattering of $K^{\pm}$ from $^{12}$C, in terms of three
$\alpha$-clusters with the parameters given in Table 1.
 The starting parameters which were taken from Ref. \cite{AAB} were used in the search involving all six parameters of the real and
  imaginary potentials, in groups of two at a time. The best fitting optical model parameters are listed in Table 1.  The central optical potential
used here is obtained from (Eq. (9)) and inserted into the  DWUCK4
program. The obtained results for best fit angular distributions of
differential elastic and inelastic cross sections shown in Figs.
1--2 and 4--6 are obtained with the parameters of Table 1. The
calculated $\chi^{2}$ values corresponding to each case under
consideration are minimum for our potential model at each of these
cases.

Examples are taken at momenta ranging from 635 to 800 MeV/c, where
the experimental data are available. At the very forward angle, the
experiments show that the differential cross sections rise very
rapidly, but the calculations converge toward smaller values. This
originates from the fact that we have neglected the contribution of
Coulomb interaction which is the long range interaction in the
optical potential.

In Ref. \cite{AZB} three values of the range parameter of $^{12}$C
are used namely $\mu$ = 0.02967, 0.04667, or 0.0884 {\em fm$^{-2}$},
one of them, $\mu$= 0.04667 {\em fm$^{-2}$}, was calculated through
the root mean square radius of $^{12}$C taking into account the root
mean square radius of alpha particle. Moreover, it produced better
fits to $^{^{12}}$C ions elastic scattering data from several
targets at several energies. To confirm this result, Figs. 1 and 2
show the present fitting of $K^{\pm}$-–$^{12}$C elastic scattering
at 635 to 800 MeV/c kaon lab momenta
 using the above mentioned different values of $\mu$. Figures 1 and 2
show that $\mu$ = 0.04667 {\em fm$^{-2}$} is preferred since it yields a better fitting especially at forward
angles. Therefore, we have chosen $\mu$ = 0.04667 {\em fm$^{-2}$} for the use in the present calculations. This agrees well with
  the reported results for the $\alpha$-particle model of $^{12}$C -$^{12}$C reaction
  \cite{MMD}. The resulting real and
imaginary optical potentials for $K^{+}$ scattered from $^{12}$C at
715 MeV/c are shown in Fig. 3. These potentials for $^{12}$C are
calculated with $\mu$= 0.0467, 0.02967, and 0.0884 {\em fm$^{-2}$}.
From that figure it is seen that in this case the real/imaginary
part is quite shallow and everywhere attractive/absorptive. These
relative shallow potentials for positive kaons compared to pions
confirm that K$^{+}$ mesons weakly interact with nuclei. The present
optical potential of $^{12}$C predicts well the maximum and
 minimum  positions with those calculated in Ref. \cite{LIQ} for
 the elastic and inelastic scattering differential
 cross sections at large angles according to the choice of the wave
  function with $\mu$=0.04667 {\em fm$^{-2}$}.
In particular,  our calculations predict two diffraction minima, but the
predicted minima are much deeper than those observed.

In Fig. 1,  the measured elastic scattering differential cross
sections at forward angles and the positions of minima and maxima
agree well with our calculations at the three momenta 635, 715, and
800 MeV/c.
 The differential elastic cross sections of $K^{-}$ scattered from $^{12}$C at 800 MeV/c is shown in Fig. 2. The $K^{-}$ minima
  are sharper than those for $K^{+}$, but the $K^{+}$ data fall more steeply than the $K^{-}$. The $K^{+}$ minima occur
   at larger angles (typically by about 5$^{o}$) than for $K^{-}$ indicating that $K^{+}$ sees a smaller nucleus than $K^{-}$. This agrees well with
  the reported results in Ref. \cite{MAR}.

Inelastic scattering is another important aspect of the interactions
of kaons with nuclei. The $\alpha$-particle model optical potential
may be tested to predict observables of $K^{\pm}$
inelastically-scattered from nuclei. In a phenomenological approach,
it is expected that both elastic and inelastic scattering will be
described in the same framework. This consistency condition may
provide more information on the kaon-nucleus potential, and is
essential before reliable nuclear structure information can be
reported. With the zero-range distorted wave Born approximation, we
may also
 compute inelastic scattering to the collective 2$^{+}$ and
 3$^{-}$ states of  $^{12}$C using the standard light ion code DWUCK4 \cite{PDK}.

  When the potential is deformed (expression (10)) and the value of the
 deformation length $\delta_{l}$ is to be adjusted, a
 reasonable agreement with data can be found as shown in
 Figs. 4--6. Coulomb excitation was found to be
 unimportant for the inelastic scattering considered here.
 The resulting deformation lengths determined by visually
 adjusting the calculations to reproduce the equivalent data listed
 in Table 2 compared to those obtained from calculations using
 parameters given in Refs. \cite{AAB,EBR}.\\ To calculate deformation lengths
 from parameters of Refs. \cite{AAB,EBR}, we used the relations:
\begin{eqnarray}
\delta_{r}=\beta_{l}\, r_{u}\, A^{\frac{1}{3}},\nonumber\\
\delta_{i}=\beta_{l}\, r_{w}\, A^{\frac{1}{3}},
\end{eqnarray}
where $\beta_{l}$, $R_{u}(=r_{u}\,A^{1/3})$, and $R_{w}(=r_{w}\,A^{1/3})$ are defined and given in
Ref. \cite{AAB}. The deformation lengths extracted from the present
calculation are consistent with those obtained by the analysis of
the inelastic scattering of $K^{+}$ and other particles from
$^{12}$C as shown in Table 2. In our case the corresponding
values fall between 0.802 to 1.429  {\em fm} for different momenta.

The extraction of $\delta_{2,3}$ for $K^{\pm}$ inelastic scattering
presents opportunity for investigating possible differences in the
proton and neutron distributions in $^{12}$C nucleus. No extra
neutrons in the $^{12}$C nucleus (N=Z) may lead to different values
of ${\delta^{+}}_{2,3}$ and ${\delta^{-}}_{2,3}$. The shell closure
at A=12 for the 2$^{+}$ and 3$^{-}$ states is collective. The
excitation of both proton and neutron particle-hole configurations
are equally likely to occur, and no significant difference between
$\delta^{+}$ and $\delta^{-}$ are expected for scattering to either
the 2$^{+}$ or 3$^{-}$ state in $^{12}$C.

The angular distributions for the inelastic scattering of
 $K^{+}$ by $^{12}$C (2$^{+}$; 4.44 MeV) at 635 to 800 MeV/c kaon Lab momenta are
 presented in Fig. 4. The experimental angular distributions and the present calculations
   are generally smooth, without the presence of well-defined
 minima. Rather poorer agreement is obtained for
 calculations of $K^{+}$ scattering to the strong coupling of the ground state to the 2$^{+}$ excited state in
 $^{12}$C at 715 MeV/c in large-angle region, satisfactory
agreement with the measured angular distributions is obtained at 800 MeV/c, although the general shapes of the
 angular distributions are well reproduced by the present calculations. The peak of the
cross sections versus scattering angles shifts forward with
increasing momenta.

Also shown in Fig. 5, angular distributions of the inelastic
scattering of $K^{+}$ by $^{12}$C (3$^{-}$; 9.64 MeV) at the three
momenta considered here, except for $K^{+}$ scattering to the
3$^{-}$ at 715 MeV/c. Inelastic scattering calculations with the
present potential describe the shapes of the inelastic angular
distributions better than the calculations of WS with optical
parameters fit to the measured elastic scattering data \cite{REC}.

The theoretical predictions for $K^{-}$ to the 2$^{+}$ and 3$^{-}$
excited states of $^{12}$C 800 MeV/c are found to be in better
 agreement with data, as shown in Fig. 6.  The present calculations based on the WS potential of $\alpha$ clusters of
  $^{12}$C  give a better fit
 with data than the calculations of Marlow {\it et al.} \cite{MAR}.

From Table 2, it can be seen that values of deformation lengths
determined here using the present potential are very similar to
those obtained in Refs.\cite{AAB,EBR}.
  It is clear from Table 2 that the deformation
lengths of the real potential
 are larger than the corresponding
ones for the imaginary potential in all cases under consideration.
Moreover, $\delta_{r}$ for the 2$^{+}$ and
 3$^{-}$ states at
715 MeV/c given in Table 2 falls adequately close to the
corresponding $\delta$ from different reactions given in Refs.
\cite{BOH}--\cite{SMI}. It can be seen from
 Table 2 that the
values of the imaginary deformation lengths determined here
  for
2$^{+}$ and 3$^{-}$ excited states in $^{12}$C decrease with
increasing kaon momenta, except for the case of 715 MeV/c $K^{+}$
inelastic scattering off the 3$^{-}$ state in $^{12}$C.

 The present potential is also used to predict
reaction and total cross sections of $K^{\pm}$ scattering on
$^{12}$C from 400 to 1000 MeV/c. The DWUCK4 program calculates these
cross sections. The results of the present model (solid curves) are
smaller than calculations based on the Watanabe superposition model
  (dashed  curves) \cite{EBR} by
3\% for the total and 3-5\% for the reaction cross sections. The two
computations are very similar to each other and this is due to that
both use the Woods-Saxon shape for both the real and imaginary parts
of the kaon--alpha potential (see Eq. (5) in Ref. \cite{EBR}). It
can be seen from Fig. 7 that there is a somewhat agreement between
the present calculations of $\sigma_{R}$ for $K^{+}$ and data
reported in Ref. \cite{EFR}. This indicates that the imaginary part
of the present optical potential used here, which is strongly
correlated to $\sigma_{R}$, is well predicted.   Since the kaon mean
free path $\lambda$ is proportional to the inverse of the total
$K$-nucleon cross sections i.e. $\lambda \propto
\frac{1}{\sigma_{T}/A}$ \cite{DOV}, it is easy to see from Fig. 7
that low values of $\sigma_{T}$  when compared to the corresponding
values for pion
 scattering \cite{AAE,ASA} indicate longer mean free path
for $K^{+}$.

In Table 3, reaction and total cross sections and ratios of
experimental \cite{EFR} to calculated cross sections are shown for
$^{12}$C. The results show that both calculated reaction and total
cross sections are smaller than experimental data. Calculated
reaction cross sections display larger discrepancy
 with the data than the total cross sections.  The ratios in Table 3 display a momentum dependence which can be
removed when the values for $^{12}$C are referred to those for
$^{6}$Li. These super-ratios \cite{EFR} are then nearly momentum
independent. The present calculations are in good agreement with
data at low momenta  but they go down rapidly  with the data for
higher momenta.

In general, the shapes and magnitudes for the $K^{+}-^{12}$C elastic
and inelastic cross sections are similar using the present model
 which is based on the 3$\alpha$-cluster of $^{12}$C and those of Watanabe superposition model which is also based on the $\alpha$
 clusters \cite{EBR}, although the two computations do not lead to identical total and reaction cross sections for the  $K^{+}-^{12}$C scattering
 at the momentum range considered here.

\section{Conclusion}
We used a very simplified nonrelativistic optical potential, the
present model (with real and imaginary Woods-Saxon folding
potentials) to obtain a good description for
 $K^{\pm}$-$^{12}$C elastic scattering at 635, 715, and 800
MeV/c kaon lab momenta using the distorted wave Born
approximation (DWBA) with DWUCK4 code. The success of the present potential to fit the differential elastic and inelastic cross section data may
 be due to the fact that the kaon-alpha potential is based on the Woods-Saxon form which was forced to agree with the data.

The differential inelastic scattering of $K^{\pm}$ to the lowest 2$^{+}$ and
3$^{-}$ states in $^{12}$C are calculated by the present local potential,
 and this potential has the ability to reproduce
the shape and magnitude of the $K^{\pm}$ scattering data. The
deformation lengths determined from the present analysis are in
agreement with deformations determined previously using
 other probes. This collective model analysis is
successful in its description of the reaction dynamics and the
structure of N=Z nucleus.

In conclusion,  optical model used in the present work can serve as
a reliable model for kaon-nucleus scattering.

\section{ACKNOWLEDGMENT}
I would like to thank Professor A.L. Elattar, Department of Physics,
Assiut University, for a careful reading of the manuscript.
\newpage



\newpage

\vskip 1 cm \noindent {\bf Fig. 1}  Differential elastic scattering
 cross sections for 635, 715, and 800 MeV/c $K^{+}$ on $^{12}$C.  The solid curve is the optical potential predictions from the present work with
$\mu$=0.04667 {\em fm}$^{-2}$, the dotted curve is
the same calculations but with $\mu$=0.0884 {\em fm}$^{-2}$, and the dashed curve is
the same calculations but with $\mu$=0.02967 {\em fm}$^{-2}$. The solid circles represent
 the experimental data taken  from Refs. \cite{MAR,REC}

\vskip 1 cm \noindent {\bf Fig. 2} As in Fig. 1, but for $K^{-}$ elastic scattering from $^{12}$C at 800 MeV/c. The solid circles represent
 the experimental data taken  from Ref. \cite{MAR}

\vskip 1 cm \noindent {\bf Fig. 3} Left, the real parts of the
optical potentials for $K^{+}$ scattered from  $^{12}$C at 715
MeV/c. The full curves represent the predictions of the expression
(9), with $\mu$=0.02967, 0.04667, and 0.0884 {\em fm$^{-2}$}. Right,
the same as left but for the imaginary parts.

\vskip 1 cm \noindent {\bf Fig. 4} Differential inelastic scattering
cross sections of 635, 715, and 800 MeV/c $K^{+}$
  leading to the lowest 2$^{+}$ state in $^{12}$C with the present optical potential with $\mu$=0.04667 {\em fm}$^{-2}$. The
experimental data are taken from Refs. \cite{MAR,REC}

\vskip 1 cm \noindent {\bf Fig. 5} As in Fig. 3, but for $K^{+}$
exciting the 9.64 MeV 3$^{-}$ state of $^{12}$C

\vskip 1 cm \noindent {{\bf Fig. 6}} As in Fig. 2, but for $K^{-}$
exciting the 4.44 MeV 2$^{+}$ and 9.64 MeV 3$^{-}$ excited states of
$^{12}$C

\vskip 0.5 cm \noindent {\bf Fig. 7} The top portion is for total
cross sections for $K^{-}$-$^{12}$C. The lower portion is for the
reaction and
 total cross sections for $K^{+}$-$^{12}$C. Solid lines are from the present work and dashed lines are the Watanabe model calculations of
  Ref. (Ebrahim 2011 \cite{EBR}). These lines are compared with the data from Refs. (E. Friedman, et al.,  1997  \cite{EFR}, R.A. Krauss, et al.,
   1992 \cite{KRA} and D.V. Bugg, et al., 1968 \cite{BUG}) for the total and reaction cross sections
\newpage
\noindent {\bf Table 1}
The phenomenological optical potential parameters for $\alpha$-particles, they are parameterized in a Woods-Saxon form.
 A
positive real potential is attractive and a positive imaginary
potential is absorptive
 \begin{center}
\begin{tabular}{ccccccccc}
\hline
 Reaction&P$_{_{lab}}$&$V$&$r_{v}$&$a_{v}$&$W$&$r_{w}$&$a_{w}$\\
&(MeV/c)&(MeV)&(fm)&(fm)&(MeV)&(fm)&(fm)\\
\hline
K$^{\pm}$-$\alpha$&635&56.58&1.075&0.538&45.32&0.923&0.483\\
                   &715&80.32&1.221&0.605&42.05&0.904&0.403\\
                   &800&100.25&1.025&0.502&65.71&0.916&0.342\\
 \hline
\end{tabular}
\end{center}

\noindent {\bf Table 2}
 Deformation lengths $\delta$ (in {\em fm}) from $K^{+}$ meson inelastic scattering
from $^{12}$C compared to those predicted by
others.
\begin{center}
\begin{tabular}{ c c c c| cc c|c c}
\hline
&&\multicolumn{2}{c|} {Present}&\multicolumn{5}{c} {Others}\\
$P_{lab}$&state&$\delta_{r}$&$\delta_{i}$&$\delta_{r}$&$\delta_{i}$&Ref.&$\delta$&Ref.\\
 \hline
635&$2^{+}$&1.429&1.029&1.725&1.114&\cite{AAE}&1.22--1.27($\pm$0.13)&\cite{BOH}\cr
&&&&1.522&0.991&\cite{AAB}&1.03--1.63&\cite{NAD}\cr
715&&1.214&0.921&1.355&1.005&\cite{AAE}&1.02--1.41&\cite{DEM}\cr
&&&&1.735&1.014&\cite{AAB}&1.5--2.1&\cite{MOF}\cr
800&&1.308&0.902&1.413&0.978&\cite{AAE}&1.0$\pm$0.1&\cite{PHA}\cr
&&&&1.619&1.017&\cite{AAB}&1.07$\pm$0.05&\cite{MEB}\\
635&$3^{-}$&1.232&0.878&1.449&1.102&\cite{AAE}&0.65-1.23&\cite{SMI}\cr
&&&&1.381&0.897&\cite{AAB}&&\cr
715&&1.118&0.842&1.302&1.205&\cite{AAE}&&\cr
&&&&1.432&0.866&\cite{AAB}&&\cr
800&&1.208&0.802&1.319&0.955&\cite{AAE}&&\cr
&&&&1.358&0.835&\cite{AAB}&&\cr
\hline
\end{tabular}
\end{center}

\noindent {\bf Table 3}
 Reaction and total cross sections and ratios of the experimental to calculated values for $K^{+}$ scattering from $^{12}$C.
  The experimental values are taken from \cite{EFR}.
\begin{center}
\begin{tabular}{ c c ccc }
\hline
P$_{lab}$&$\sigma_{R}$&$\sigma_{T}$&$\sigma_{R}(exp.)/\sigma_{R}(cal.)$& $\sigma_{T}(exp.)/\sigma_{T}(cal.)$\\
\hline
450&116.2&156.4&&\\
488&118.2&159.6&1.019&1.017\\
500&122.2&160.1&&\\
531&126.3&162.8&1.024&1.023\\
600&133.3&166.5&&\\
656&138.6&169.5&1.023&1.031\\
700&143.1&170.8&&\\
714&144.6&171.5&1.033&1.024\\
800&153.2&176.8&&\\
\hline
\end{tabular}
\end{center}

\end{document}